\documentstyle[epsf]{aipproc}

\begin{document}

\title {A New Cosmological Paradigm: the Cosmological Constant and Dark Matter}

\author{Lawrence M. Krauss\footnote{Material based on Invited Lectures, 
PASCOS 98, Boston MA,  March 98; Tropical
Workshop on Particle Physics and Cosmology, San Juan, Puerto Rico, April 98; 
WEIN98, Santa Fe, NM, June 98. To appear in these proceedings. Research supported
in part by the DOE. email:krauss@theory1.phys.cwru.edu}}
\address{
Departments of Physics and Astronomy \\
Case Western Reserve University, 
Cleveland, OH~~44106-7079
}

\maketitle
\begin{abstract}
\noindent The Standard Cosmological Model of the 1980's is no more.  I
describe the definitive evidence that the density of matter is insufficient to
result in a flat universe, as well as the mounting evidence that the
cosmological constant is not zero.  I finally discuss the implications of these
results for particle physics and direct searches for non-baryonic dark matter,
and demonstrate that the new news is good news.
\end{abstract}

\section{Introduction}

One of the great developments of the 1980's was the creation of a Standard Model
of Cosmology based on ideas arising from Particle Theory.  This model involved
the following trilogy of ideas:
\vskip 0.2in

\begin{center}
\begin{tabular}{lc}
(1) & $\Omega \equiv 1$ \\
& \\
(2) & $\Lambda \equiv 0$ \\
& \\
(3) & $\Omega_{\rm matter} \approx \Omega_{\rm CDM_{\rm axion}^{\rm WIMP}} \ge
0.9$ \\
\end{tabular}
\end{center}

\vskip 0.3in

A decade later observational cosmology has made tremendous strides, and we now
know that at least two of these fundamental notions must be incorrect.  Either

\vskip 0.2in

\begin{center}
\begin{tabular}{lc}
(1) & $\Omega \ne 1$ \\
& \\
(2) & $\Lambda \equiv 0$ \\
& \\
(3) & $\Omega_{\rm matter} \approx \Omega_{\rm CDM_{\rm axion}^{\rm WIMP}}
\approx  0.1 - 0.3$ \\
\end{tabular}
\end{center}

\vskip 0.2in

or

\vskip 0.2in

\begin{center}
\begin{tabular}{lc}
(1) & $\Omega \equiv 1$ \\
& \\
(2) & $\Lambda \ne 0$ \\
& \\
(3) & $\Omega_{\rm matter} \approx \Omega_{\rm CDM_{\rm axion}^{\rm WIMP}}
\approx 0.1-0.3$ \\
\end{tabular}
\end{center}

\vskip 0.2in

In either case the implications for both cosmology and particle physics are
profound.  In the first place, 

\vskip 0.2in

\begin{center}
{\bf Either: $\Omega \ne 1$  or  $\Lambda \ne 0$ }
\end{center}

\vskip 0.1in

Whichever is true, this implies we don't understand something very fundamental
about the microphysics of the Universe---a very exciting prospect!  If $\Omega
\ne 1$ then the canonical prediction of inflation is incorrect, and we have to
understand how inflation, or another theory, might address the fine tuning
required to solve the flatness problem without actually resulting in a flat
universe today.  If $
\Lambda \ne 0$ then the situation is in a sense even more exciting, as there is
no theory of the cosmological constant at the present time, and the supposition
that this quantity is indeed zero rests primarily on {\it a priori} theoretical
prejudice at this point.  (I here include in the term "cosmological constant"
those models which involve a very slowly varying scalar field, which in effect
mimics a cosmological constant over long time intervals.)

At the same time, we have:

\vskip 0.2in

\begin{center}
\begin{tabular}{lc}
& $\Omega_{\rm matter} \approx \Omega_{\rm CDM_{\rm axion}^{\rm WIMP}}
\approx 0.1-0.3$ \\
\end{tabular}
\end{center}

\vskip 0.2in

This also has dramatic implications, not only for our understanding of the role
dark matter plays in the formation of large scale structure, but also for our
propsects for direct detection of non baryonic dark matter.  Contrary to one's
naive expectations however, the implications are quite positive.  Dark
matter may not contribute $90 \%$ of the mass of the Universe, as previously
envisaged,
but it still appears to outweigh baryonic matter.  Moreover, as I will
demonstrate, in all cases these results suggest that the interaction strength of
dark matter with normal matter will be {\it INCREASED}, and thus in principle
direct detection should be { \it easier} than it would otherwise be.   As long
as the dark matter contribution to the fraction of the closure density is
larger than 0.1, so that it can account for essentially the entire inferred
dark matter content of galactic halos in general, and our galactic halo in
particular, the increase in interaction cross section is not counterbalanced by
a decrease in the dark matter flux on earth, so that the net event rate in
detectors should be larger than would be the case if $\Omega_{\rm CDM} \approx
0.9$.

\section{The Case for a Cosmological Constant}

Over the past 5 years a variety of indirect observables, involving the three
fundamental independent observables in cosmology, the expansion rate, the
matter content, and large scale structure, have all suggested that either the
universe is open, or the cosmological constant is not zero
\cite{kraussturner,ostrikerstein,krauss}.  In the past year, the indirect
evidence has been strengthened by new large scale structure measurements, and
for the first time, striking new direct measurements suggest that the Hubble
expansion is accelerating.   I first review the older, indirect evidence, and
then describe the most recent results.

\subsection{The Age Problem}

The Hubble constant, by a very simple argument, gives an upper limit on the age
of a matter dominated universe.   Matter causes a deceleration of the universal
expansion over time.  Thus, at earlier times the universe would have been
expanding faster than it is at the present time.  One can, in turn, therefore
derive an upper limit on the age of the universe by considering the fact that
all galaxies were once located together, and using the relation for
a constant velocity to determine the length of time a galaxy at a given 
distance, moving away at a constant velocity took to get there, i.e.
$d = vt \rightarrow t=d/v = H_0^{-1}$, where the definition of the Hubble
constant, $H_0$ has been used.   In fact, of course, this upper limit on the
age of the Universe is an overestimate of its age, and with a given
cosmological model one can derive a specific relation between the Hubble
constant today and the age of the universe.  One has:

\begin{tabular}{lr}

Flat matter dominated &  $t =(2/3) H_0^{-1} = 9.7 {\rm Gyr} (65/H_0)$ \\
& \\
Open ($\Omega >0.2$) & $t < .85 H_0^{-1} = 12.5 {\rm Gyr} (65/H_0)$ \\
& \\
Flat ($\Omega_{\Lambda} < 0.8$) & $t < 1.08 H_0^{-1} = 16 {\rm Gyr} (65/H_0)$ \\
\end{tabular}

Thus, if one could definitively demonstrate that the Universe were older than
either of the first two relations allowed, given the allowed range of $H_0$, one would
have strong evidence that $\Lambda \ne 0$, since a non-zero cosmological
constant would allow a universal {\it acceleration}, and hence allows an older
universe for a fixed Hubble Constant.

While precisely such a situation seemed to prevail as recently as 1996
\cite{chab}, more recent estimates for the age of globular clusters have
suggested that the age of our galaxy is younger than previously estimated. At
the same time, estimates of the Hubble constant are now somewhat lower than
previously claimed, so that a range of 65- 75 kms$^{-1}$Mpc$^{-1}$ is now
preferred \cite{Freedman}.  Nevertheless, the new quoted 95
$\%$ lower limit, of approximately 9.8 Gyr with a best estimate of the age of
11.5 Gyr
\cite{chab2}, strongly disfavors a flat universe, even if it remains compatible
with an open universe.

\subsection{The Baryon Problem}

Big Bang Nucleosynthesis (BBN) has for some time provided an upper limit on the
total density of baryonic matter in the univere \cite{krausskern,copist}:

\begin{equation}
\Omega_B h^2 \le .026
\end{equation}

Most recently, claimed observations of the deuterium fraction in primordial
hydrogen clouds illuminated by the light of distant quasars \cite{tytler}
suggest that the actual baryon abundance is near the upper limit of this range.
While this puts pressure on BBN analyses, more germaine for this discussion is
the fact that when combined with direct observations of the baryon fraction on
large scales today, it effectively rules out the possibility of a flat universe.

X-Ray Observations of Clusters of Galaxies, the largest bound structures known in
the Universe suggest that the dominant baryonic material in these systems exists
in the form of hot X-Ray emitting gas.  Assuming this material is in hydrostatic
equilibrium with the gravitational potential of these systems one can, by
observing both the X-Ray Luminosity and temperature as a function of radius,
perform an inversion which gives an estimate of this potential, and hence the
total mass, $M_T$,  of these .  At the same time, direct observations of the
luminosity yield an estimate for the total baryonic mass in hot gas, and hence
the total baryonic mass $M_B$.  Thus, one derives the ratio $ R = M_B/M_T$ for
these sytems.  Now, as these systems are the largest bound objects known, it is
reasonable to assume that they are good probes of the distribution of all
gravitating matter on large scales.  Thus, the ratio R is expected to be not
just the ratio of baryon to total mass of clusters, but the ratio of baryonic
to total mass in the Universe.  Thus,{\it if} the Universe is flat, so that the
density corresponding to $M_T$ yields $\Omega =1$ then one has precisely the
relation $R = \Omega_B$.   Therein lies the problem. Observations, combined with
theoretical models of clusters yield the constraint \cite{evrard}

\begin{equation}
R > .043 h^{-3/2}
\end{equation}

If $R = \Omega_B$ then clearly this equation is inconsistent with the BBN
bound.  

This problem can be simply resolved however, if $\Omega_{M_T} < 1$ so that the
ratio R is in fact larger than $\Omega_B$.   There are then two possibilities.
Either  $\Omega_{M_T} = \Omega <1$, or $\Omega_{M_T} + \Omega_{\Lambda} =1$.

\subsection{Large Scale Structure}

The growth of structure in the Universe, if gravity is responsible for such
growth, provides an excellent probe of the universal mass density, based largely
on issues associated with causality alone.   The basic idea is the following: 
If primordial density fluctuations have no preferred scale, then one can express
their Fourier transform as a simple power of the wavenumber $k$.  At the same
time, if this power is much greater than unity, density fluctuations will blow up
for large wavenumber, or small wavelength, and too many primordial black holes
will be created.  If the power is much less than unity, then fluctuations on
large scales (small wavenumbers) will be inconsistent with the observed isotropy
of the Cosmic Microwave Background radiation.   Thus, we expect the exponent,
$n$ to be near one, and inflationary models happen to predict precisely this
behavior.  

The primordial power spectrum, however, is not what we observe today, as
density fluctuations can be affected by causal microphysical processes once the
scale of these fluctuations is inside the horizon scale---the distance over
which light can have travelled between t=0 and the time in question.  One can
show that in an expanding universe, as long as the dominant form of energy
resides in radiation, gravity is ineffective at causing the growth of density
fluctuations.  In fact, such primordial fluctuations in baryons will be damped
out due to their coupling to the radiation gas.   Once the universe becomes
matter dominated, however, primordial fluctuations on scales smaller than the
horizon size can begin to grow.

\begin{figure}[htb]
\epsfxsize =\hsize \epsfbox{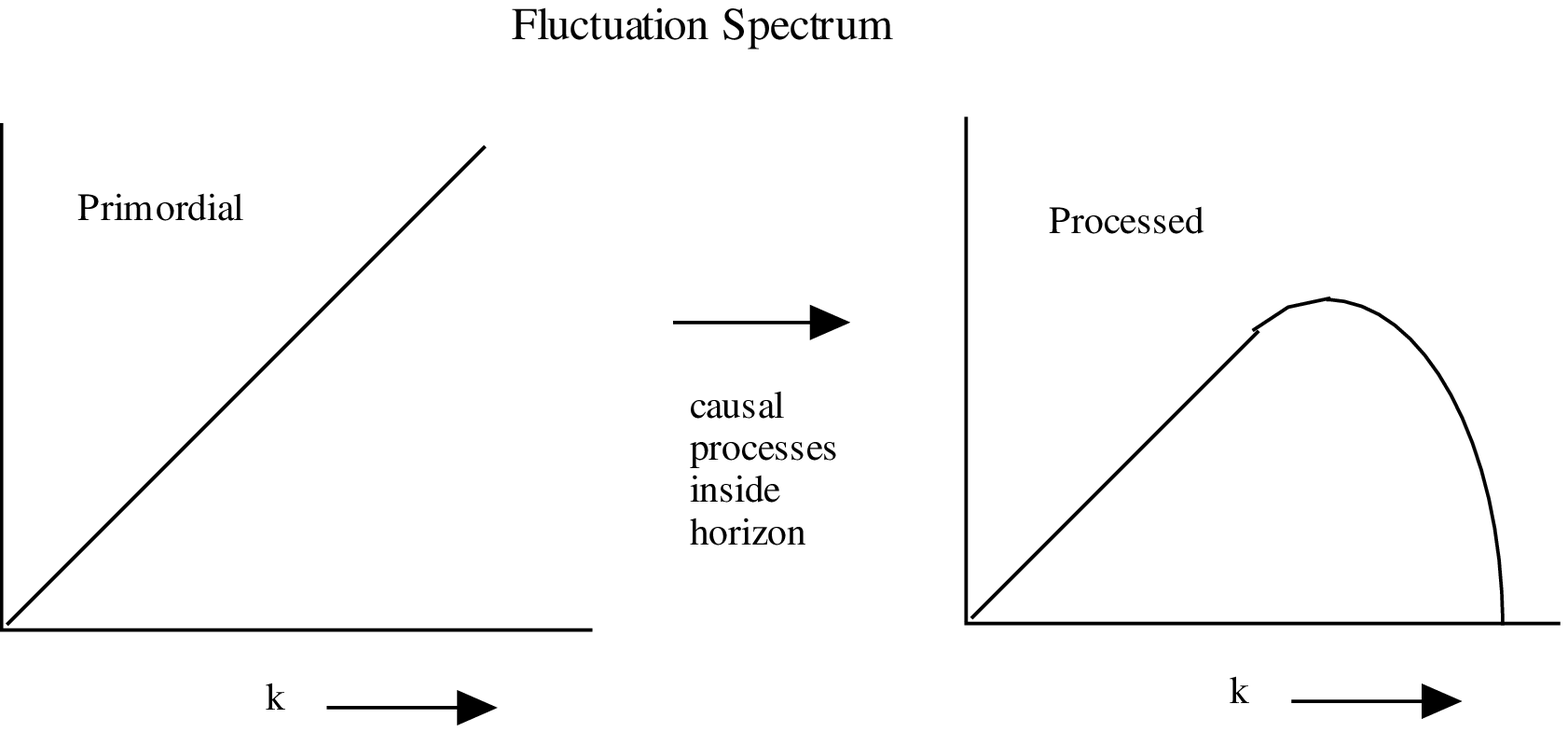}
\end{figure}

These arguments suggest that an initial power law spectrum of fluctuations will
``turn over" as shown
in Figure 1 above 
for large wavenumbers which entered inside the causal horizon during
the early period of radiation domination in the Universe.  By exploring the
nature of the clustering of galaxies today over different scales, including
measurements of the two point correlation function of galaxies, the angular
correlation of galaxies across the sky on different scales, etc, one can hope to
probe the location of this turn-around, and from that probe the time, and
thus the scale which first entered the horizon when the universe became matter
dominated.  Clearly this time will depend upon the ratio of matter to radiation
in the Universe today (if this ratio is increased, then matter, whose density
decreases at a slower rate than radiation as the universe expands, will begin to
dominate the expansion at an earlier time, and vice versa. In turn, knowing this
ratio today gives us a handle on $\Omega_{\rm matter}$. A recent compilation of
large scale galaxy clustering data \cite{peacockdodds,pd2} 
restricts this quantity to be in the range:

\begin{equation}
0.25 \le \Omega_{\rm matter}h \le 0.35
\end{equation}

Since $h$ appears to lie in the range 0.65-0.75, this suggests $\Omega_{\rm
matter} <1$.   Note, however that this argument does not restrict the component
of $\Omega$ which might reside in a cosmological constant today, since this
energy density is fixed, so that even if it dominates today, it was irrelevant
compared to the energy density of matter and radiation at early times.

\vskip 0.1in

By combining these three independent sets of constraints, one can, for either an
open universe, or a flat universe with a cosmological constant, constrain the
parameter space of $h$ versus $\Omega_{\rm matter}$ \cite{kraussturner,krauss}.
These constraints are shown in the figures below \cite{krauss}, 
which clearly indicate that a
flat matter dominated universe appears to be inconsistent with observations.

\begin{figure}[htb]
\begin{minipage}[htb]{62mm}
\epsfxsize = 2.0in \epsfbox{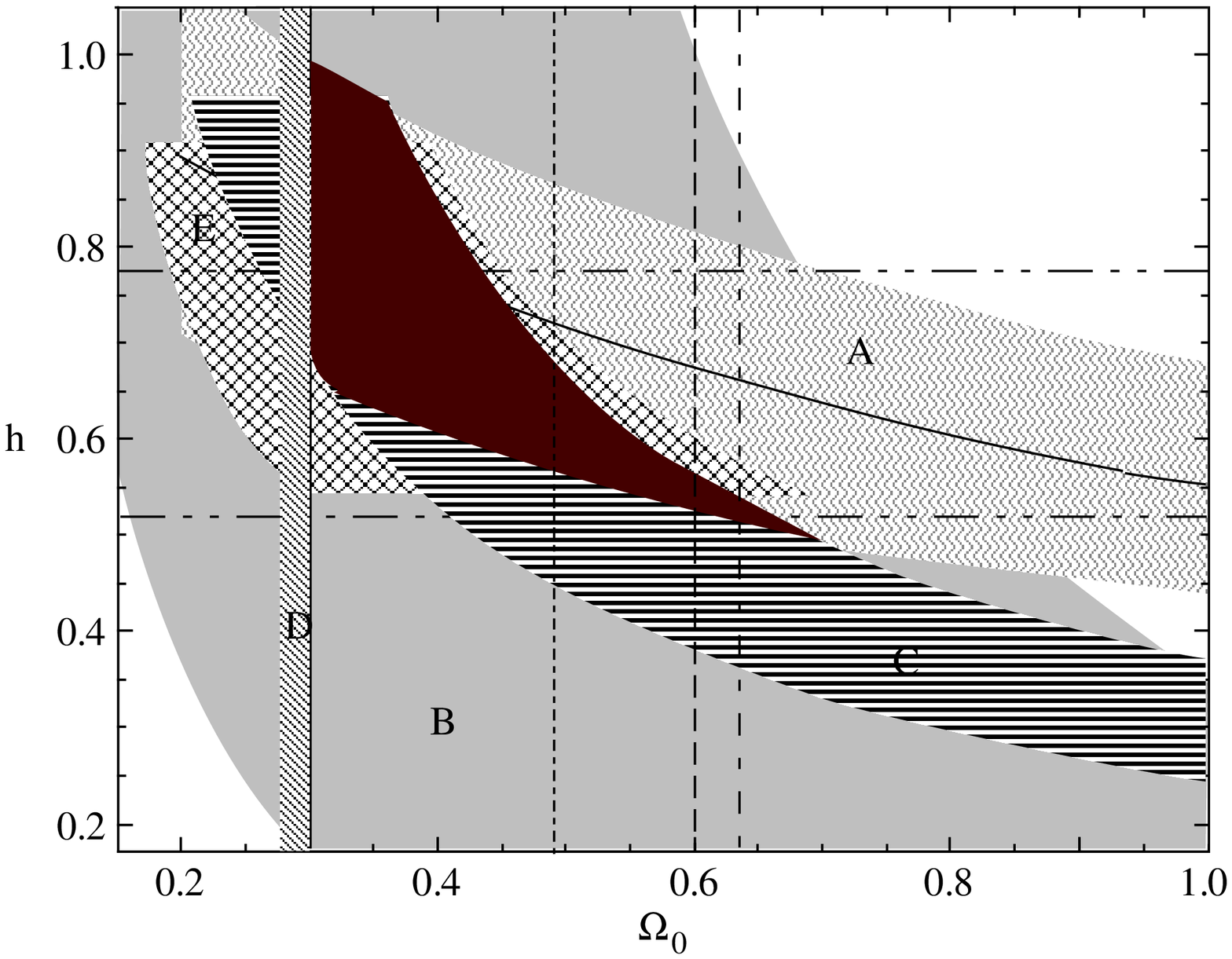}
\caption{Constraints on $h$ vs $\Omega_{\rm mat}$ for
a flat universe. The
constraints discussed in the text (A, B, and C). The
other constraints are discussed in (Krauss 98).}
\end{minipage}
\hspace{\fill}
\begin{minipage}[htb]{62mm}
\vskip -5mm
\epsfxsize = 2.25in \epsfbox{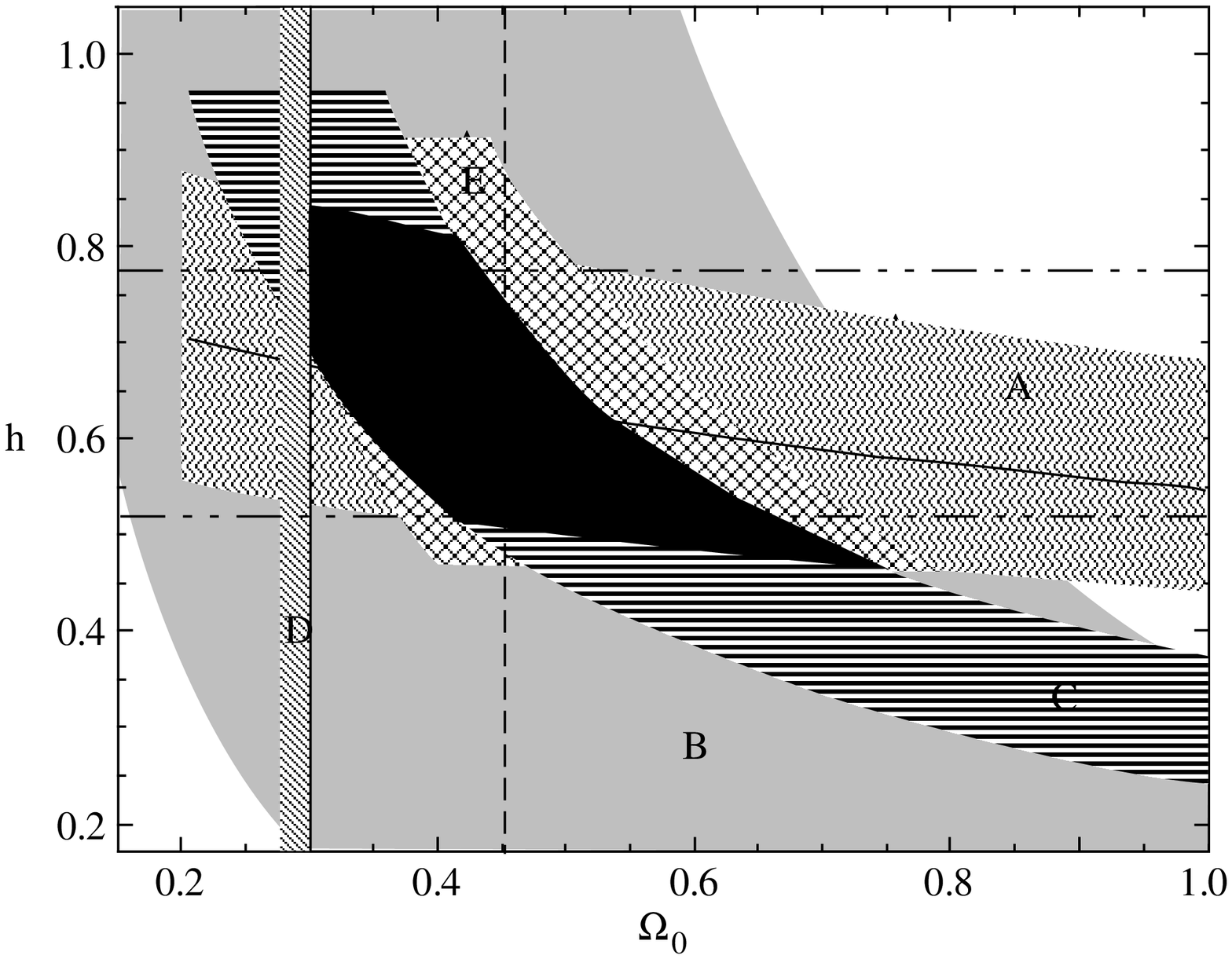}
\caption{Constraints on $h$ vs $\Omega_{\rm matter}$ for
an open universe. Constraints are as described in previous figure.}
\end{minipage}
\end{figure}

\subsection{Recent Observations: Galaxy Clustering, the CMB and Supernova
Standard Candles}

The situation described above has been recognized for over two years.  Within
the last year or so, however, new observations have made the case for a
cosmological constant even stronger.  I briefly describe these observations
below. 

\vskip 0.1in

(1) Cluster Evolution:  In a Universe in which there is not a precisely critical
density of matter, small density fluctuations on large scales cease to grow due
to gravity once either the total density begins to deviate significantly from
that associated with a flat universe, or when the density in a cosmological term
begins to dominate over the density of matter.  Thus, if the universe is not
flat today, or if the cosmological constant dominates, then structure has ceased
to continue to grow on large scales.  If, however, the universe is flat and
matter dominated, structures on every larger scales are continuing to form by
gravitational collapse.   This suggests that if one examines out to high
redshifts one should see significantly fewer rich clusters of galaxies then one
sees today.  The difference is significant.  As Neta Bahcall and colleagues have
recently pointed out \cite{bahcall}, the probability of finding a rich cluster at a
redshift of 0.7 is perhaps 100 times smaller for a flat matter dominated
universe than for a universe in which the growth of large structures stopped
some time ago.  A single large cluster observed at such high redshift can then
provide, largely independent of detailed modelling, damning evidence against a
flat matter dominated universe.  By cataloguing such clusters out to redshifts
of this order, these authors have recently claimed to present definitive
evidence ruling out a flat, cold dark matter dominated universe.  Both open,
or flat cosmological constant dominated universes are consistent with this data.

(2) Type 1a Supernova at High Redshift:  As anyone who has glanced at a paper in
the past six months knows, two groups have recently and hopefully independently 
claimed to measure the relation between redshift and distance out to redshifts 
in excess of 0.5, and in so doing have been able to probe for cosmic deceleration
or acceleration.  The probes used have been Type 1a Supernovae.  These have been
claimed to be superb Standard Candles for two reasons: (i) Type 1a supernovae
occur when a white dwarf, through accretion, passes the Chandrsekhar limit, and
undergoes a detonation explosion.  The physics of this process should not
depend significantly on the evolutionary status of the galaxy in which the
star is housed. (ii) Detailed studies of the luminosity profile of such 
supernovae suggest a strong relation between the width of the light curve, and
the absolute luminosity of the supernova.  This allows one, in principle, to
accurately determine this absolute luminosity.   Based on these features both
groups have now claimed to report definitive evidence for a non-zero cosmological
constant.  Moreover, they claim, at the 99 $\%$ confidence level, to be able
to rule out both a flat, and an open universe with zero cosmological constant 
\cite{perl,kirsh}.
Even more remarkably, the favored region, for a flat universe, is precisely in
the range favored by the other constraints in the figures shown above (which I
remind you were drawn before these results appeared), namely
$ \Omega_{\rm mat} \approx 0.3-0.4, \  \Omega_{\Lambda} \approx 0.6-0.7$.  It
remains to be seen if further data taken at high redshift confirm these results, 
and more importantly confirm the assumption that evolution is negligible for
such supernovae.

(3) CMB preliminary studies:  If one decomposes the observed CMB anistropies
on the sky into multipoles on the sky, it is well known that CDM cosmologies
predict a rise in the power spectrum as a function of multipole approaching 
a large peak, followed by smaller peaks.  The position, in multipole space, of
this first peak is a probe of the geometry of the universe, as it is related to
the angular size of the horizon at the last scattering surface, as seen today.
{\it Very} preliminary results from terrestrial observations of high multipole 
anisotropies in the CMB tend to confirm the existence of such a peak, and moreover
the position of the peak appears to favor a flat, versus an open universe.  If
this is the case, then the existence of a non-zero cosmological constant, in
light of all the other direct and indirect evidence, seems assured.

\vskip 0.1in

It is almost unnerving that all existing cosmological data appears to
point in the same direction---towards a non-zero cosmological constant. As
someone who has been promoting the idea that the cosmological constant might
be non-zero for some time, I frankly found myself more comfortable when some
of the data argued against this possibility.  In any case, the wealth of
cosmological data now available appears to unambiguously point to
the fact that $\Omega_{\rm mat} < 1$, whether or not the cosmological constant
is non-zero.  This fact may have profound implications for dark matter detection.

\section{Implications for Particle Physics and the Search for Dark Matter}

The magnitude of the 
cosmological constant which would be required by the present data is
remarkably small. Before proceeding to examine the consequences of the above
results for dark matter detection, it is worth pausing for a moment to reflect
on this feature.  When I recently did this while preparing this lecture,  
I turned to a pocket calendar I had with me, and noticed the quotation:

\begin{center}
{\it ``To see what is in front of one's nose requires a constant struggle''}
\rightline{George Orwell}
\end{center}

What, you may ask, does this have to do with the topic at hand.  Plenty, I claim.
For it reminds us that we can put remarkably stringent limits on certain 
quantities by using macroscopic amounts of material.  In particular, it harkens
back to another famous quotation, this time from Maurice Goldhaber, who put
one of the first limits on proton decay by declaring that if the proton had
a lifetime less than about $10^{17}$ years, ``{\it You could feel it in your
bones!''}.  By this he meant that proton decays in our body would be so
frequent that we would die from the radiation exposure.

In this spirit we can perform a similar experiment.   Look at the end of your
nose.   Now, in a universe dominated by a cosmological constant, space begins
to expand exponentially.  One can calculate than for distances separated by
larger than an amount $R > M_{Pl}/3 \Lambda^{1/2}$, points will have a relative
velocity exceeding that of light, and thus will remain out of causal contact.
Thus, the fact that you can see the end of your nose implies a bound
$\Lambda < 10^{-68} M_{Pl}^4$!

Of course, the fact that we can see distant galaxies gives us an even stronger
bound.  And, the fact that the cosmological constant affects dynamics on larger
scales no more than it is claimed to by the present observations gives a bound
$\Lambda < 10^{-123} M_{Pl}^4$.   What makes this small number so hard to 
understand, in a cosmological context is not merely the ``naturalness'' problem
of which particle physicists are aware, but rather, if this has been constant
over cosmological time, this is the first time in the history of the universe
when the energy density in a cosmological constant is comparable to the
energy density of matter and radiation!  It is for this reason that some
cosmologists are driven to the idea that what is being observed is not really
a cosmological constant, but something perhaps more exotic \cite{stein}.

Be that as it may.  Particle physicists will never measure a quantity of this
magnitude directly in the laboratory.  However, they may one day directly measure
non-baryonic particles which presumably make up our galactic halo.  And the
new data brings good tidings in this regard.   For the only well motivated 
candidates for Cold Dark Matter, axions and WIMPS one can write down a general
relation:

\begin{equation}
\sigma_{detection} \approx {1 \over \Omega_{DM}}
\end{equation}

The reasons for this are different for each candidate.  For axions, one can
understand the origin of this relation as follows:  Axions are dark matter
because at early times thier potential (considered as a function of an 
angular variable which can be taken to go from $-\pi$ to $\pi$) changes form:
\begin{figure}[h]
\epsfxsize = \hsize \epsfbox{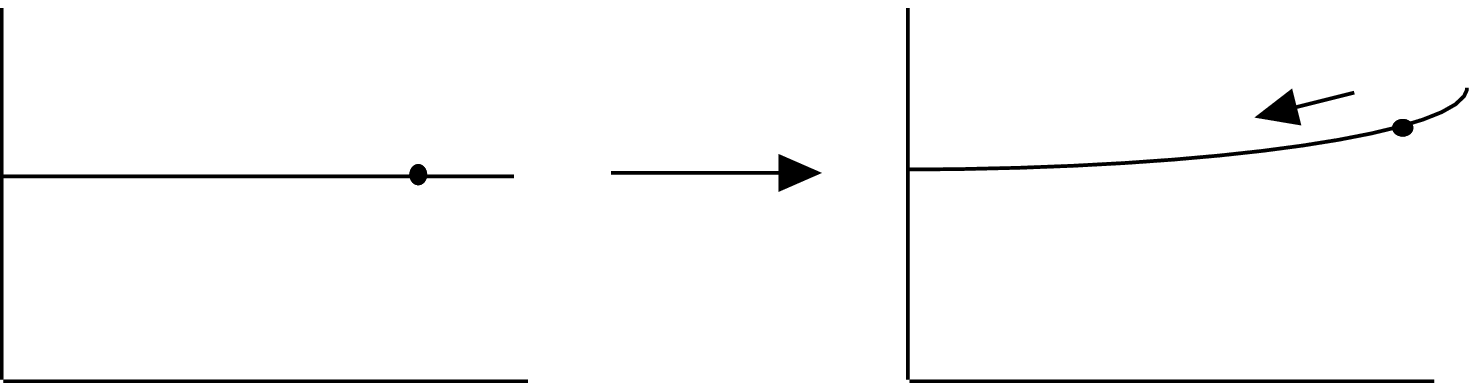}
\end{figure}

In the former case, no energy is stored in the axion field.  However, 
once the axion gets a mass, energy is stored in the axion field, which
then dynamically rolls to the bottom of its potential.  However, the
time it takes to begin rolling is inversely proportional to the 
curvature of its potential, and is thus inversely proportional to the
axion mass.  Thus, the smaller the axion mass, the longer the energy
gets stored before it begins to redshift and the greater the remnant axion
density.  Since the axion couplings are inversely proportional to the
axion mass, one therefore obtains the relation above.

For WIMPs, the situation is more direct. Remnant WIMPs results from
incomplete annihilations in an initial thermal poulation, so that 
\begin{equation}
\Omega_X h^2 \approx { 10^{-37} {\rm cm^2} \over <\sigma_{ann} v>}
\end{equation}

By crossing symmetry, the WIMP annihilation cross section is roughly
proportional to the WIMP scattering cross section.  Thus, as the WIMP
abundance decreases, its scattering cross section generally increases.

Astute experimentalists may argue that this is a scam, because as the
WIMP (axion) density decreases, the flux on Earth also decreases, so
even if there are larger cross sections, the event rate will not change!
However, this is wrong.  Until the density decreases to the point (below
about $\Omega_x < 0.1$ ) when WIMPs (axions) do not have sufficient
densities to account for all galactic halo dark matter, it is natural to
assume that their galactic density is given by the halo density.  Just
because their overall cosmic density is insufficient to close the universe,
this need not imply that their flux on earth is reduced!
 
\section{Conclusions} 

It is time to throw in the towel and accept the paradigm shift in 
Cosmology.  All evidence suggests that $\Omega_{DM} < 1$.  The dominant
energy density in the universe may be far darker than that stored in
dark matter ---it may be stored in empty space itself!  Nevertheless
the news is good for direct detection of non-baryonic dark matter.
Cross sections may be higher than previously invisaged when it was
felt that Cold Dark Matter must result in a closure density all by itself.

Of course, as a theorist one tries to think beyond the next set of experiments.
What if the next generation of WIMP detectors detects a signal, for
example?  What then?  How will we be sure that it is from the galactic halo,
and how can we learn about the halo properties, and/or the properties of
the dark matter particles?  I will close this lecture by advertising some
new work we have been involved in which may shed some light on new
WIMP signatures.  First, by exploring the angular variation of the predicted
WIMP signals which might arise from a variety of different models for our
galactic halo, we have recently demonstrated \cite{copiheokrau} that
as few as 15-20 events would be needed in a detector having directional
sensitivity before a halo induced signal could be distinguished from a
flat background of noise.  Next, with T. Damour, I have recently demonstrated
that there is likely to be a new solar system population of WIMPs existing
in trapped Solar orbits intersecting the earth if the WIMP cross sections on
matter are large enough to be detected at the next generation of detectors 
\cite{damkrauss}.  This population will produce a dramatically different 
signal in cryogenic detectors, and could be used as a discriminant to
verify any previoulsy detected WIMP signal, or could be searched for 
independently.

I thank my collaborators involved in various aspects of the work
described here, Michael Turner, Brian Chaboyer, Pierre Demarque, Peter Kernan,
Craig Copi, Junseong Heo, and Thibault Damour.
I also thank the those who organized a very stimulating and enjoyable meeting.


\begin{thebibliography}  {999}
\bibitem{kraussturner} L.M. Krauss and M. S. Turner, J. Gen. Rel. Grav., {\bf 27},1137 
(1995)
\bibitem{ostrikerstein} J.P. Ostriker and P. Steinhardt, Nature, {\bf 377}, 600
(1995)
\bibitem{krauss} L.M. Krauss, Ap. J., {\bf 501}, in press, (1998)
\bibitem{chab} B. Chaboyer, P. Demarque, P. Kernan, and L. M. Krauss,
Science, {\bf 271}, 957 (1996)
\bibitem{Freedman} W. L. Freedman, Proc. Nat. Acad. Sci., {\bf 95}, 2 
(1998)
\bibitem{chab2}B. Chaboyer, P. Demarque, P. Kernan, and L. M. Krauss,
Ap. J. , {\bf 494}, 96 (1998)
\bibitem{krausskern} L. M. Krauss and P. J. Kernan, Phys. Lett. {\bf B347}, 347 
(1995)
\bibitem{copist} C. Copi, D.N. Schramm, and M.S. Turner, Science, {\bf 267}, 192
(1995)
\bibitem{tytler} D. Tytler, X. M. Fan, and S. Burles, Nature, {\bf 381}, 207 
(1996)
\bibitem{evrard} A. E. Evrard, MNRAS, in press (1998)
\bibitem{peacockdodds} J.A. Peacock and S.J. Dodds, MNRAS, {\bf 267}, 1020 (1994)
\bibitem{pd2} A. R. Liddle, D.H. Lyth, P.T.P. Viana, M. White, MNRAS, {\bf 282},
281 (1996)
\bibitem{bahcall} N.A. Bahcall, X. Fan, Ap. J. {\bf 504}, in press, (1998); 
Proc. Nat. Acad. Sci., {\bf 95}, 5956 (1998)
\bibitem{perl} S. Perlmutter {\it et al}, LBNL preprint 41801 (1998)
\bibitem{kirsh} A. Reiss {\it et al}, preprint, submitted to Ap. J.
\bibitem{stein} R. Caldwell {\it et al}, Phys. Rev. Lett., {\bf 80}, 1582 (1998)
\bibitem{copiheokrau} L.M. Krauss, Phys. Rep. (Proc. Workshop on Dark Matter
Detection), in press; see also C. Copi, J. Heo and L.M.Krauss, in preparation.
\bibitem{damkrauss} T. Damour and L.M. Krauss, astro-ph/9806165 (1998)
\end{thebibliography}
\end{document}